\pdfoutput=1
\def\repropreprint{}
\ifdefined\repropreprint
  \documentclass[sigconf,nonacm,balance=false]{acmart}
\else
  \documentclass[sigconf,pbalance]{acmart}
\fi
\usepackage{amsmath}
\usepackage{algorithmic}
\usepackage{xurl}
\usepackage{tabularx}
\usepackage{tikz}
\usepackage[local]{gitinfo2}
\AtBeginDocument{%
  \fancypagestyle{versionedstandard}[standardpagestyle]{%
    \fancyfoot[L]{\scriptsize\ttfamily Source version: \gitDescribe}%
  }%
  \fancypagestyle{versionedfirst}[firstpagestyle]{%
    \fancyfoot[L]{\scriptsize\ttfamily Source version: \gitDescribe}%
  }%
}
\usetikzlibrary{arrows.meta,positioning}

\ifdefined\repropreprint
  \AtBeginDocument{\raggedbottom}
\else
\setcopyright{rightsretained}
\acmDOI{}
\acmISBN{}
\acmConference[REPRO-HPC 2026]{1st Workshop in Sustainable
Practices for Reproducibility in HPC}{June 2026}{Hamburg, Germany}
\acmYear{2026}
\copyrightyear{2026}
\acmBooktitle{Publia Open Conferences Insights, Volume 1: Proceedings of 1st Workshop in Sustainable Practices for Reproducibility in HPC (REPRO-HPC)}
\editor{Quentin Guilloteau}
\editor{Valérie Hayot-Sasson}
\editor{Dennis Hoppe}
\editor{Josef Weidendorfer}
\fi

\begin{document}
\title[Layered Reproducibility for HPC Applications]{Layered Reproducibility for High-Performance Computing Applications: The Feel++ and Ktirio Urban Buildings Case Study}
\ifdefined\repropreprint
  \subtitle{Preprint --- initial submitted version}
\fi

\author{Javier Cladellas}
\email{javier.cladellas@cemosis.fr}
\affiliation{%
  \institution{Cemosis, IRMA UMR 7501, University of Strasbourg, CNRS}
  \city{Strasbourg}
  \country{France}
}

\author{Vincent Chabannes}
\email{vincent.chabannes@cemosis.fr}
\affiliation{%
  \institution{Cemosis, IRMA UMR 7501, University of Strasbourg, CNRS}
  \city{Strasbourg}
  \country{France}
}

\author{Christophe Prud'homme}
\email{christophe.prudhomme@cemosis.fr}
\affiliation{%
  \institution{Cemosis, IRMA UMR 7501, University of Strasbourg, CNRS}
  \city{Strasbourg}
  \country{France}
}

\renewcommand{\shortauthors}{Cladellas et al.}

\begin{abstract}
The High-Performance Computing (HPC) community struggles with reproducibility due to complex software stacks, heterogeneous hardware, and expensive validation cycles. This contribution presents a layered reproducibility strategy in which Feel++ provides a traceable software supply chain and Ktirio Urban Buildings (KUB) builds on it to manage versioned urban-building data, executable simulations, and checksum-verifiable result artifacts. Feel++ treats continuous integration (CI) plans, CMake presets, packaging manifests, public software metadata, operations commands, containers, software bill of materials (SBOM)/provenance metadata, and Feel++ Benchmarking reports for Feel++ and KUB cases as repository-owned artifacts. KUB applies this foundation to City Energy Model Database (CEMDB) datasets, manifests, Findable, Accessible, Interoperable, and Reusable (FAIR) packaging, and verifiable outputs for computationally expensive simulations. The case study shifts reproducibility from late manual packaging to an artifact chain linking source commits, packages, containers, benchmark configurations, dataset versions, provenance records, and generated results.
\end{abstract}


\maketitle
\pagestyle{versionedstandard}
\thispagestyle{versionedfirst}

\section{Introduction}\label{sec:introduction}
The HiDALGO2 project focuses on accurate and fast simulations addressing global challenges. Within this framework, the Ktirio Urban Buildings (KUB) pilot models building energy dynamics from a building scale to a city scale through City Energy Model (CEM) simulations. KUB depends on Feel++, a C++ library for solving a large range of partial differential equations using Galerkin methods \cite{prudhomme2012feelpp}. Feel++ is available as open-source software at \url{https://github.com/feelpp/feelpp} and archived on Zenodo \cite{prudhomme2026feelppzenodo}. KUB reproducibility is therefore inseparable from Feel++ reproducibility: Feel++ continuous integration (CI), operations tooling, package metadata, public software metadata, and container-generation logic determine which numerical software stack can be rebuilt, audited, and reused by the pilot.

Deploying such complex applications across diverse supercomputing environments frequently leads to inconsistencies and dependency conflicts. To efficiently simulate city-scale domains, the underlying computational meshes and associated data must be partitioned and distributed across multiple nodes. Consequently, subtle variations in Message Passing Interface (MPI) implementations, hardware interconnects, and partitioning libraries across different clusters severely complicate portability, acting as a major barrier to scientific progress.

Current high-performance computing (HPC) reproducibility practice combines several strands. Reproducible computational research emphasizes recording code, parameters, data, and execution context \cite{sandve2013reproducible}; artifact packaging tools such as ReproZip capture runnable experiments after the fact \cite{chirigati2016reprozip}; HPC software environments rely on package managers, containers, and system-aware testing to control portability and performance variability \cite{gamblin2015spack,merkel2014docker,kurtzer2017singularity,karakasis2020reframe}; and Findable, Accessible, Interoperable, and Reusable (FAIR), software bill of materials (SBOM), and provenance practices address data reuse and software supply-chain transparency \cite{wilkinson2016fair,ntia2021sbom,torresarias2019intoto}. This paper is positioned at the intersection of these strands: it connects repository-owned software supply-chain records, continuous correctness and performance benchmarking, and KUB data/run evidence so that minimal reproducible experiments and expensive simulation artifacts are inspectable.

To address these methodological and technical challenges, this contribution treats sustainable reproducibility as a layered artifact chain integrated into the software lifecycle. Reproducibility here includes not only rebuilding and rerunning the software, but also reproducing and comparing performance behavior across European High-Performance Computing (EuroHPC) systems; Feel++ Benchmarking, available as open-source software at \url{https://github.com/feelpp/benchmarking} and archived on Zenodo \cite{cladellas2025feelppbenchmarking}, provides this performance-reproducibility layer for both Feel++ and KUB cases. The contributions are: (i) a Feel++ foundation based on repository-owned continuous integration and continuous deployment (CI/CD), build presets, operations commands, packages, images, performance benchmarking, and public metadata; (ii) a KUB data and execution layer using City Energy Model Database (CEMDB) layouts, manifests, schemas, and checksums for urban-building inputs and outputs; and (iii) a stakeholder-oriented path for users, clients, partners, decision makers, and technical evaluators linking source commits, packages, containers, benchmark configurations, performance reports, dataset versions, provenance records, and generated results.

Section~\ref{sec:strategy} presents the layered reproducibility strategy: Feel++ provides the repository-owned software supply chain through CI/CD, CMake presets, operations tooling, public metadata validation, package generation, Open Container Initiative (OCI)/Apptainer artifacts, and SBOM/provenance preparation; Feel++ Benchmarking \cite{cladellas2025feelppbenchmarking} then supports continuous correctness and performance benchmarking for both Feel++ and KUB cases. Section~\ref{sec:evidence} focuses on the KUB data and execution evidence, describing the CEMDB layout, versioned datasets, run manifests, provenance records, schemas, checksums, and artifacts used to make expensive simulations inspectable without requiring full reruns.

\section{Layered Reproducibility Strategy}\label{sec:strategy}

Achieving sustainable reproducibility requires separating the generic numerical software supply chain from the domain-specific application and data record. In our strategy, Feel++ provides the reusable HPC foundation and KUB consumes that foundation to execute city-energy simulations and preserve their evidence. This keeps the software environment, benchmark definition, dataset layout, and run record connected without making KUB responsible for rebuilding the entire numerical stack.

\subsection{Feel++ Reproducibility Foundation}

The software stack required for Feel++ and KUB relies on Boost, Gmsh, PETSc, SLEPc, Eigen, ParMetis, MPI, and related libraries. Host-specific modules and local build recipes are fragile because compiler versions, operating-system updates, interconnect drivers, and partitioning libraries vary across clusters. Feel++ therefore treats the execution environment as a repository-derived artifact: OCI images \cite{merkel2014docker} are built from declared package and image metadata, tested in CI, and executed on HPC systems through Apptainer \cite{kurtzer2017singularity}.

The Feel++ development repository makes CI and packaging state part of the source tree so that package definitions, image recipes, and metadata are pinned to the software source they build. The target catalog is declared in \path{.github/plan-ci.json}; reproducible build and test workflows are encoded in \path{CMakePresets.json}; packaged components are declared in \path{packaging/manifest/components.toml}; and \texttt{ops/} commands generate packages, Spack environments, Dev Container profiles, Docker Bake contexts, and version data. In particular, \texttt{fpp-pkg}, \texttt{fpp-spack}, \texttt{fpp-version}, and \texttt{fpp-dev} resolve package versions, inspect repository-owned Spack environments, generate development containers, and prepare image build contexts; package workflows upload artifacts and publish Debian/Ubuntu repository snapshots through the Feel++ Advanced Package Tool (APT) infrastructure. The public metadata CLI \texttt{fpp-ip}, where IP denotes public intellectual-property metadata, records software identity, license signals, component and dependency inventories, Docker and Apptainer inventories, release metadata, deterministic YAML/JavaScript Object Notation (JSON) exports, and source metrics. This metadata strategy supports Software Package Data Exchange (SPDX) 3.0.1 and CycloneDX SBOM exports, together with provenance records for Docker and Apptainer artifacts identified by immutable digests or file hashes.

\subsection{Feel++ Benchmarking and Continuous Benchmarking}

Feel++ Benchmarking addresses continuous benchmarking by consuming the built software artifacts rather than redefining the environment, and it applies to both core Feel++ benchmarks and KUB CEM cases. The \texttt{feelpp.benchmarking} repository implements a Python layer around ReFrame-HPC \cite{karakasis2020reframe}, is available at \url{https://github.com/feelpp/benchmarking}, and is archived on Zenodo \cite{cladellas2025feelppbenchmarking}. A benchmark run is described by JSON-controlled machine, benchmark, and report configurations whose schemas validate the executable, timeout, resource request, platform, input files, optional Girder dependencies, parameter sweeps, sanity rules, scalability files, and report contents. At runtime, \texttt{feelpp-benchmarking-exec} pulls the declared Apptainer image when required, downloads declared remote inputs, invokes ReFrame, and writes \texttt{reframe\_report.json}; \texttt{feelpp-benchmarking-render} turns reports into an Antora dashboard \cite{antoraDocs}. Machine configurations for EuroHPC systems such as Karolina, MeluXina, Discoverer, and Vega separate scheduler, module, launcher, and container details from the benchmark itself.

The same tool therefore makes performance reproducibility measurable over time. For Feel++, it records whether reusable numerical components, packages, containers, and launcher configurations remain executable and comparable across systems. For KUB, it binds the same runtime artifacts to CEM-specific benchmark descriptions, declared input data, resource requests, sanity checks, performance records, and reports. The workflow can run benchmarks directly or through Slurm, upload reports to Girder, and stage dashboard updates through auditable pull requests. Feel++ Benchmarking is thus used to monitor performance evolution, detect regressions, and compare execution behavior across EuroHPC systems for both Feel++ and KUB.

\subsection{KUB Application and Data Layer}

KUB is the application layer built on top of the Feel++ foundation. Its CI/CD workflows build and test the Python package and C++ CEM executable, generate Docker and Apptainer artifacts, publish package artifacts, and run CEM simulations that upload result directories. In practice, the main workflow builds the wheel, runs Python tests, builds the C++ code, packages Debian artifacts, and builds container images; the CEM simulation workflow downloads declared remote data, runs \texttt{feelpp\_kub\_cem}, uses \texttt{feelppdb} as part of the simulation run output, uploads the resulting CEMDB location results, and summarizes the produced manifest. KUB therefore uses Feel++ executables and containers, but adds domain-specific reproducibility records: versioned CEMDB location datasets, preprocessing caches, single and ensemble run directories, setup summaries, provenance records, and hashed Hierarchical Data Format 5 (HDF5)/Parquet outputs.

\begin{figure}[!t]
\centering
\begin{tikzpicture}[
    font=\scriptsize,
    node distance=0.32cm,
    flowstep/.style={
        draw,
        rounded corners=1pt,
        align=center,
        text width=0.82\columnwidth,
        minimum height=0.48cm,
        inner sep=2.5pt
    },
    artifact/.style={
        flowstep,
        fill=black!5
    },
    arrow/.style={
        -{Latex[length=1.7mm]},
        line width=0.35pt
    }
]
\node[flowstep] (source) {Source and metadata\\Commit, tag, CI target, public metadata};
\node[flowstep, below=of source] (validate) {Build and test\\CMake presets, Python/C++ tests, packages};
\node[artifact, below=of validate] (package) {Runtime artifacts\\Debian/Ubuntu, Spack, OCI, Apptainer, SBOM/provenance};
\node[flowstep, below=of package] (deploy) {EuroHPC execution\\System-specific scheduler, launcher, container settings};
\node[flowstep, below=of deploy] (benchmark) {Feel++ Benchmarking\\Feel++/KUB configs, ReFrame reports, dashboard};
\node[artifact, below=of benchmark] (data) {KUB evidence\\CEMDB versions, manifests, HDF5/Parquet, hashes};

\draw[arrow] (source) -- (validate);
\draw[arrow] (validate) -- (package);
\draw[arrow] (package) -- (deploy);
\draw[arrow] (deploy) -- (benchmark);
\draw[arrow] (benchmark) -- (data);
\end{tikzpicture}
\Description{Six connected stages link source and metadata to build and test, runtime artifacts, EuroHPC execution, Feel++ Benchmarking, and KUB evidence.}
\caption{Layered reproducibility workflow: Feel++ supplies the software artifact chain, while KUB adds domain data and run evidence.}
\label{fig:workflow}
\end{figure}
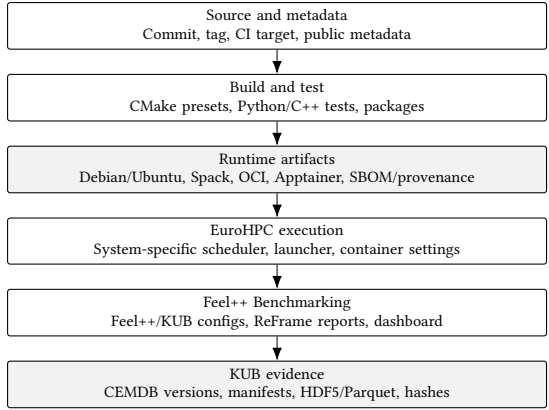

This split gives stakeholders a concrete path through the artifact chain. The Feel++ layer answers which numerical software, package set, and image were used; the shared \texttt{feelpp-benchmarking} layer records the Feel++ or KUB benchmark configuration and report; and the KUB layer answers which city dataset, preprocessing state, simulation setup, output files, schemas, and checksums were produced. Fig.~\ref{fig:workflow} summarizes this dependency from source-code changes to benchmark and CEMDB evidence.

The result is not a single container recipe but a traceable relation between source commits, CI targets, package and image identifiers, public metadata, benchmark reports, CEMDB dataset versions, and run manifests.

\begin{table}[!t]
\caption{Stakeholder-oriented artifact path.}
\label{tab:artifact-path}
\centering
\footnotesize
\begin{tabularx}{\columnwidth}{@{}>{\raggedright\arraybackslash}p{0.24\columnwidth}>{\raggedright\arraybackslash}X@{}}
\toprule
\textbf{Step} & \textbf{Evidence} \\
\midrule
Source state & Git tag or commit, \path{.github/plan-ci.json}, \path{CMakePresets.json}, public metadata export \\
Software artifact & Debian/Ubuntu package, Spack environment, OCI image, Apptainer image file, digest or hash \\
Benchmark check & Feel++ or KUB \texttt{feelpp-benchmarking} JSON, \texttt{reframe\_report.json}, dashboard entry \\
KUB dataset & \path{cemdb/locations/<id>/versions.json}, \path{v<version>/manifest.json}, DVC or remote source metadata \\
KUB run & \path{simulations/.../manifest.json}, setup summary, provenance record \\
Result validation & HDF5/Parquet artifacts, schema identifiers, SHA-256 and composite hashes \\
\bottomrule
\end{tabularx}
\end{table}

\section{KUB Data Management and Run Evidence}\label{sec:evidence}

Software traceability is only useful for KUB if the urban data and generated results are also described precisely. KUB therefore uses CEMDB as the source of record for location datasets, preprocessing caches, simulator resources, and simulation runs. In the C++ code, the \texttt{PathManager} class is the single path authority and separates shared simulator resources from location data, standalone runs, ensemble runs, and per-member ensemble outputs. In the Python layer, dataset commands generate, pack, synchronize, version, and validate CEMDB manifests.

\subsection{CEMDB Layout and Dataset Versions}

A location dataset is organized under \path{cemdb/locations/<id>/}. The \path{versions.json} file selects the active dataset version, while \path{v<version>/manifest.json} records the dataset name, schema version, KUB compatibility, source metadata, directories, contents, and checksums. The versioned directory contains the concrete inputs used by CEM runs: \path{geo/} for geographic information system (GIS) data and meshes, \path{weather/} and \path{air-quality/} for time-dependent environmental data, \path{scenarios/} for schedules and building-use mappings, and \path{preprocessing/} for partitioned static inputs. Large location archives can be referenced through Data Version Control (DVC) pointers \cite{dvcDocs} to remote stores, while the local CEMDB layout remains the execution contract.

\subsection{Run Records and Verifiable Outputs}

Each execution is isolated under \path{simulations/single/<run_id>/} or \path{simulations/ensemble/<run_id>/}. The run directory contains a manifest, setup summary, provenance record, and database artifacts. The manifest records the run id, run type, status, artifact base, referenced files, and generated outputs. The provenance record captures the Feel++ and KUB versions, command line, Git commit, branch and dirty state, MPI world size, backend, host, and Coordinated Universal Time (UTC) timestamps.

Generated outputs are not described only by filenames. HDF5 artifacts store building metadata and time series, while Parquet artifacts store global and building-level key performance indicator (KPI) summaries. Manifest entries attach schema identifiers, schema versions, SHA-256 hashes, and, for partitioned outputs, per-file hashes plus a composite hash. The Python manifest validator checks referenced files and hashes, and strict validation requires schema and checksum metadata. This lets stakeholders inspect or validate the result package without relying on the original HPC filesystem.

\subsection{Artifacts for Expensive Simulations}

City-scale and ensemble simulations are too expensive to rerun routinely for every evaluation. KUB therefore preserves reusable intermediate and summary artifacts: preprocessing stores partitioned GIS, mesh, weather, scenario, and building metadata inputs; simulator resources are isolated from run outputs; and ensemble/statistics code writes lightweight Parquet KPI summaries for downstream analysis. Minimal \texttt{feelpp-benchmarking} cases can still exercise the executable and container environment, while CEMDB manifests and hashes preserve the evidence for larger runs.

This creates two reproduction modes. The minimal mode reruns selected Feel++ or KUB benchmark cases to check that the tagged source state, package or image, EuroHPC execution configuration, remote inputs, executable, and report generation still work together \cite{berti2025ktirio}. The full-evidence mode is used for expensive city-scale and ensemble simulations: instead of rerunning every case during evaluation, it validates the archived CEMDB run directory through manifests, provenance records, schemas, and hashes, with targeted reruns only when deeper verification is required.

\section{Conclusion}

This paper presents a layered reproducibility strategy for the HiDALGO2 KUB pilot. Feel++ supplies the software artifact chain: CI plans, CMake presets, operations commands, packages, containers, public metadata, SBOM/provenance preparation, and Apptainer deployment. Feel++ Benchmarking records executable and performance evidence for Feel++ and KUB cases, while KUB adds CEMDB datasets, preprocessing state, manifests, provenance, and hashed HDF5/Parquet outputs.

The artifact under evaluation is therefore more than a container image: it relates source commits, software packages, runtime images, benchmark reports, dataset versions, run manifests, and checksums. Stakeholders can inspect package or image identifiers, run a minimal Feel++ or KUB \texttt{feelpp-benchmarking} case, and validate KUB outputs through manifest and hash records. Larger city-scale runs remain inspectable through preprocessing caches, KPI summaries, provenance, and checksums.

The main lesson is that HPC reproducibility cannot be reduced to containerization alone. Containers constrain the software environment, but scheduler policy, filesystems, MPI stacks, datasets, and execution cost remain part of the problem. The next step is to extend this artifact chain to hybrid CPU/GPU and exascale execution, where accelerator drivers, GPU runtimes, topology, scheduler placement, memory hierarchy, container bind mounts, and MPI/GPU communication paths also affect correctness and performance. The strategy therefore needs accelerator-aware metadata for container runtime options, device visibility, node topology, compiler/runtime versions, and performance counters.

\begin{acks}
Funded by the European Union. This work has received funding from the European High-Performance Computing Joint Undertaking (EuroHPC JU) and Poland, Germany, Spain, Hungary, France, and Greece under grant agreement number 101093457. This publication expresses the opinions of the authors and not necessarily those of the EuroHPC JU and Associated Countries, which are not responsible for any use of the information contained in this publication.

Part of this work was also funded by the France 2030 NumPEx Exa-MA (ANR-22-EXNU-0002) and Exa-DI (ANR-22-EXNU-0006) projects managed by the French National Research Agency (ANR).

We thank Cemosis colleagues Gwennol{\'e} Chapron, Philippe Pin{\c c}on, Juliette Antonczak, and Feel++ and KUB stakeholders for the many discussions that shaped this work.

We acknowledge EuroHPC Development Access grants EHPC-DEV-2025D08-020, EHPC-DEV-2024D05-025, and EHPC-DEV-2023D08-047 on LUMI, MeluXina, Karolina, Discoverer, Vega, and Leonardo.

\end{acks}

\bibliographystyle{ACM-Reference-Format}
\bibliography{references,local_references}

\end{document}